\documentclass[a4paper,11pt]{article}
\usepackage{jcappub}

\usepackage{xcolor}
\usepackage{graphicx}
\graphicspath{{./figLink/}{./fig/}}
\usepackage{amsmath,amssymb,amsfonts,mathrsfs,bm}
\usepackage[%
colorlinks=true,
linkcolor=blue,
citecolor=blue,
]{hyperref}

\newcommand{\Msun}{M_{\odot}}
\newcommand{\yr}{\mathrm{yr}}

\newcommand{\pc}{\mathrm{pc}}
\newcommand{\kpc}{\mathrm{kpc}}
\newcommand{\Mpc}{\mathrm{Mpc}}

\newcommand{\Hz}{\mathrm{Hz}}

\newcommand{\kHz}{\mathrm{kHz}}

\newcommand{\rv}{\bm{r}}
\newcommand{\rvo}{\bm{r}_{O}}

\newcommand{\rtld}{\tilde{r}}
\newcommand{\gb}{g^{(\mathrm{B})}}
\newcommand{\Rb}{R^{(\mathrm{B})}}

\newcommand{\hS}[1]{S^{(#1)}}
\newcommand{\hA}[1]{A^{(#1)}}
\newcommand{\pt}{\partial_{t}}
\newcommand{\ptau}{\partial_{\tau}}
\newcommand{\pr}{\partial_{r}}
\newcommand{\Psihat}{\hat{\Psi}}
\newcommand{\xiv}{\bm{\xi}}
\newcommand{\xis}{\xi_{*}}
\newcommand{\phiL}{\phi^{L}}
\newcommand{\phitld}{\tilde{\phi}}
\newcommand{\phitldot}{\tilde{\phi}_{\tilde{\omega}}^{L}}
\newcommand{\omegatld}{\tilde{\omega}}
\newcommand{\xv}{\bm{x}}
\newcommand{\yv}{\bm{y}}
\newcommand{\ts}{t_{*}}
\newcommand{\tE}{t_{E}}
\newcommand{\tauE}{\tau_{E}}
\newcommand{\tauL}{\tau_{L}}

\newcommand{\Htld}{\tilde{H}}

\newcommand{\omegaL}{\omega_{L}}
\newcommand{\wL}{w_{L}}

\newcommand{\pFq}[2]{{~}_{#1}F_{#2}}

\newcommand{\vt}{\bm{v}_{\perp}}

\newcommand{\xivL}{\bm{\xi}_{L}}
\newcommand{\xivLb}{\bar{\bm{\xi}}_{L}}
\newcommand{\bv}{\bm{b}}
\newcommand{\tL}{t_{L}}

\newcommand{\intff}{\int_{-\infty}^{\infty}}
\newcommand{\wh}{\hat{w}}

\newcommand{\omegahat}{\hat{\omega}}

\newcommand{\tc}{t_{c}}

\newcommand{\Mc}{M_{c}}
\newcommand{\ML}{M_{L}}
\newcommand{\RE}{R_{\mathrm{E}}}

\newcommand{\avL}[1]{\bm{a}_{L_{#1}}}
\newcommand{\xivLi}{\bm{\xi}_{L_{1}}}
\newcommand{\xivLii}{\bm{\xi}_{L_{2}}}
\newcommand{\xvLi}{\bm{x}_{L_{1}}}
\newcommand{\xvLii}{\bm{x}_{L_{2}}}

\newcommand{\Psitot}{\Psi_{\mathrm{tot}}}

\newcommand{\peta}{\partial_{\eta}}
\newcommand{\ptf}{\partial_{\tf}}
\newcommand{\zetav}{\bm{\zeta}}
\newcommand{\omegaZL}{\omega_{Z_L}}
\newcommand{\tf}{\mathfrak{t}}
\newcommand{\zf}{\mathfrak{z}}

\begin{document}

\title{%
    Modulations of Gravitational Waves due to Non-static Gravitational Lenses
}

\author[a]{Xing-Yu Yang,}
\author[b,c]{Tan Chen,}
\author[d,b,e]{Rong-Gen Cai}

\affiliation[a]{Quantum Universe Center (QUC), Korea Institute for Advanced Study, Seoul 02455, Republic of Korea}
\affiliation[b]{CAS Key Laboratory of Theoretical Physics, Institute of Theoretical Physics, Chinese Academy of Sciences, Beijing 100190, China}
\affiliation[c]{School of Physical Sciences, University of Chinese Academy of Sciences, Beijing 100049, China}
\affiliation[d]{Institute of Fundamental Physics and Quantum  Technology, Ningbo University, Ningbo, 315211, China}
\affiliation[e]{School of Fundamental Physics and Mathematical Sciences, Hangzhou Institute for Advanced Study, University of Chinese Academy of Sciences, Hangzhou 310024, China}

\emailAdd{xingyuyang@kias.re.kr}
\emailAdd{chentan@itp.ac.cn}
\emailAdd{cairg@itp.ac.cn}

\abstract{%
    Gravitational waves (GWs) offer a new observational window into the universe, providing insights into compact objects and cosmic structures. Gravitational lensing, commonly studied in electromagnetic waves, also affects GWs, introducing magnification, time delays, and multiple images. While existing studies focus on static lenses, many astrophysical lenses are dynamic, with time-varying mass distributions such as moving stars or orbiting binaries. We develop a general theoretical framework to describe non-static lenses and demonstrate how they modulate GW signals, inducing unique time-varying amplitude modulations and spectral broadening. By examining uniformly moving and orbiting binary lenses, we show that these modulations provide new observational signatures, enhancing our understanding of lensing objects and the dynamics of the universe. Our findings have important implications for GW astronomy, offering novel ways to probe lens dynamics and improve the interpretations of GW signals.
}

\maketitle

\section{Introduction}

The discovery of gravitational waves (GWs) by LIGO and Virgo has revolutionized our understanding of the universe, providing a new observational channel that complements traditional electromagnetic observations~\cite{LIGOScientific:2016aoc}.
GWs are ripples in spacetime generated by cataclysmic astrophysical events, such as the mergers of black holes or neutron stars~\cite{LIGOScientific:2016sjg, LIGOScientific:2017vwq, LIGOScientific:2018mvr, LIGOScientific:2020ibl, KAGRA:2021vkt}.
Since their first detection in 2015, GWs have become a powerful tool for probing the properties of compact objects, testing general relativity, and exploring the dynamics of the cosmos~\cite{LIGOScientific:2016lio, LIGOScientific:2018cki, LIGOScientific:2018jsj}.

Gravitational lensing, a cornerstone of general relativity, occurs when a massive object between a distant source and an observer bends the path of light or GWs~\cite{1992grle.book.....S}.
This phenomenon, traditionally studied in the context of electromagnetic waves, has been increasingly explored for its effects on GWs~\cite{Nakamura:1997sw, Nakamura:1999uwi, Takahashi:2003ix, Itoh:2009iy, Sereno:2010dr, Cao:2014oaa, Lai:2018rto, Liao:2019aqq, Diego:2019rzc, Liu:2020par, LIGOScientific:2021izm, Bulashenko:2021fes, Caliskan:2022hbu, Tambalo:2022wlm, Savastano:2022jjv, Lin:2023ccz, Savastano:2023spl, GilChoi:2023ahp, Brando:2024inp, Villarrubia-Rojo:2024xcj}.
Gravitational lensing of GWs can magnify signals, create multiple images, and introduce characteristic time delays, offering new opportunities to study both the source of GWs and the lensing object itself.
Such effects can provide insights into matter distribution, including dark matter, and enhance our understanding of the large-scale structure of the universe~\cite{Dai:2016igl, Jung:2017flg}.

However, the existing literature on gravitational lensing of GWs predominantly focuses on static gravitational lenses, where the mass distribution of the lens remains constant over the timescale of the GW event.
In reality, many astrophysical lenses, such as moving stars, orbiting galaxies, or evolving galaxy clusters, exhibit time-varying mass distributions.
These non-static gravitational lenses can introduce additional complexities into the lensing process, leading to modulations in the observed GW signals that are not present in static lens models. The dynamic nature of these lenses can produce unique observational signatures, which can serve as novel probes of both the lensing objects and GW sources.

In this paper, we investigate the effects of non-static gravitational lenses on GWs, addressing a significant gap in the current understanding of GW lensing.
We develop a theoretical framework to describe time-dependent lensing effects and propose methodologies for detecting and analyzing GWs modulated by non-static lenses.
Specifically, we explore two typical examples: uniformly moving lenses and orbiting binary lenses, each presenting distinct modulating effects on GW signals.

The study of non-static gravitational lenses not only enhances our ability to interpret GW observations but also opens new avenues for studying the dynamics of lensing objects.
The time-dependent amplification factors derived in this work reveal how evolving mass distributions can alter GW waveforms, potentially leading to new observational tests of general relativity and the properties of the sources.
The non-static gravitational lensing can provide us with unique insights into dynamic and transient phenomena of the universe and has the potential to uncover the hidden nature of cosmic structures.
By accounting for the temporal dynamics of lenses, we provide a more complete picture of GW lensing and set the stage for future observational and theoretical advancements in this emerging field.

This paper is organized as follows. In Section~\ref{sec:general}, we establish the general theoretical framework for non-static gravitational lenses, including the derivation of key equations describing the time-dependent amplification factor. Section~\ref{sec:typical} presents typical examples of non-static gravitational lenses, focusing on uniformly moving lenses and orbiting binary lenses, illustrating how their dynamics affect GW signals. In Section~\ref{sec:modulation}, we analyze the modulation effects on both monochromatic and polychromatic GWs, providing a detailed discussion of the resulting spectral broadening and time-varying amplitude modulations. Finally, Section~\ref{sec:conclusion} concludes the paper by summarizing our findings and discussing potential implications for future GW observations and theoretical studies.
We set $G = c=1$ throughout the paper for notational brevity.

\section{General framework for non-static gravitational lenses}\label{sec:general}

\subsection{Wave optics}

We consider the propagation of GWs through the gravitational potential of non-static lensing objects.
The background metric is given by
\begin{equation}\label{eq:metric}
    ds^{2} = -[1+2U(t,\rv)]dt^{2} + [1-2U(t,\rv)]d\rv^{2} \equiv \gb_{ab} dx^{a} dx^{b} ,
\end{equation}
where $|U(t,\rv)| \ll 1$
is the gravitational potential of the non-static lensing objects.
The GWs $h_{ab}$ is described by the propagation equation~\cite{1973grav.book.....M}
\begin{equation}\label{eq:propagation}
    h_{ab;c}{}^{;c}+2\Rb_{cadb}h^{cd}=0 ,
\end{equation}
subject to the transverse traceless Lorentz gauge condition
\begin{equation}\label{eq:gague}
    h_{ab}{}^{;b} = 0 = h^{c}{}_{c},
\end{equation}
where the semicolon denotes the covariant derivative with respect to background metric $\gb_{ab}$, and $\Rb_{cadb}$ is the background Riemann tensor.

In the case of gravitational lensing, the GWs appear, relative to the observers of interest, as nearly plane and monochromatic on a scale large compared with a typical wavelength, but very small compared with the typical radius of curvature of spacetime, $\gb_{ab,c}\sim \mathscr{R}^{-1} \ll \lambda^{-1}$.
Following the eikonal expansion, one has~\cite{1973grav.book.....M}
\begin{equation}\label{eq:h_expansion}
    h_{ab} = \Re \left\{ e^{iS/\epsilon} \left[ A_{ab}+\epsilon B_{ab} + \mathcal{O}(\epsilon^{2}) \right ] \right\} ,
\end{equation}
where $\Re$ indicates that the real part of the expression is to be taken, and $\epsilon$ is a formal expansion parameter with eventual value unity which serves to identify orders of magnitude.
Define the scalar amplitude $A$, wave vector $k_{a}$, and polarization tensor $e_{ab}$ as
\begin{align}
    A &\equiv \left( \frac{1}{2} A_{ab}^{*} A^{ab} \right)^{1/2} ,\\
    k_{a} &\equiv S_{,a} ,\\
    e_{ab} &\equiv A_{ab}/A,
\end{align}
with $A_{ab}^{*}$ denoting the complex conjugate of $A_{ab}$.
The polarization tensor $e_{ab}$ is parallel-transported along the null geodesic, and its change due to gravitational lensing is of the order of $U$, which is very small in the observational situation~\cite{1973grav.book.....M, Takahashi:2003ix}.
Therefore, one can regard the polarization tensor as a constant, and then the propagation of GWs $h_{ab}$ can be treated as the propagation of scalar waves $\phi$, with the relation of
\begin{equation}
    h_{ab} = \Re \left\{ \phi e_{ab} \right\} \simeq \Re \left\{ A e^{iS} e_{ab} \right\} .
\end{equation}

\begin{figure}[htbp]
    \centering
    \includegraphics[]{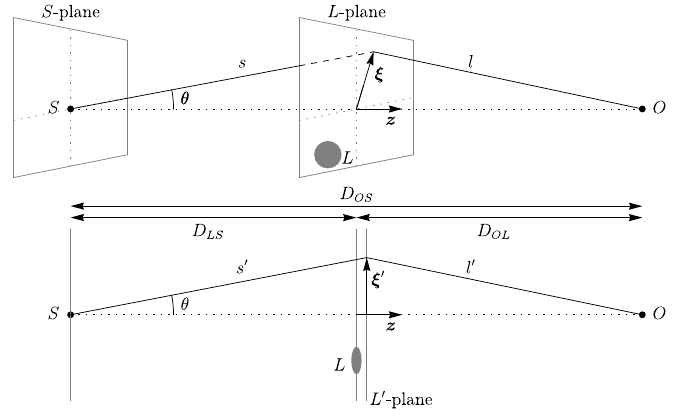}
    \caption{Diagram of the gravitational lensing system.}
    \label{fig:diagram}
\end{figure}

We consider non-static gravitational lenses whose radial motion is negligible compared with their distances to the observer and source, which is satisfied by most lens systems.
In order to solve the propagation equation with gravitational lenses, one can choose an intermediate plane $L'$ parallel to lens plane $L$
such that: (a) $L'$ is close to $L$ in the sense that the distance between $L$ and $L'$ is much smaller than the distance between $L$ and observer $O$; (b) between the source $S$ and $L'$ there are no caustics; (c) between $L'$ and observer $O$, the gravitational potential $U$ of the lens is negligible, so that there the metric may be taken to be flat~\cite{1992grle.book.....S}.
The diagram of the lensing system is shown in Fig.~\ref{fig:diagram}.

Consider the spherical monochromatic outgoing waves of frequency $\omega$ from the source, the waves have the form
\begin{equation}
    \phi(t,\rv) = \frac{1}{r} e^{-i\omega(t-r)},
\end{equation}
when there are no gravitational lenses.
Here the spherical coordinates $(r,\theta,\varphi)$ with origin at the source are used.

Because of condition (b), one can obtain the lensed field $\phiL$ in the $L'$-plane by inserting the eikonal expansion into the propagation equation and gauge condition, which gives
\begin{subequations}\label{eq:SA}
    \begin{align}
        0 &= k_{a}k^{a} ,\\
        0 &= 2A_{,a}k^{a} + A k^{a}{}_{;a} .
    \end{align}
\end{subequations}
Since $|U|\ll 1$, one can expand $\gb_{ab}$, $S$ and $A$ based on the order of $|U|$, and solve Eq.~\eqref{eq:SA} order by order.
The $\mathcal{O}(U^{0})$ term gives
\begin{subequations}
    \begin{align}
        0 &= -\pt\hS0 \pt\hS0 + \nabla\hS0 \cdot \nabla\hS0 ,\\
        0 &= -2\pt \hA0 \pt\hS0 + 2 \nabla\hA0 \cdot \nabla\hS0 + \hA0\Box\hS0 ,
    \end{align}
\end{subequations}
where $\Box \equiv -\pt^{2}+\nabla^{2}$ is the flat space d'Alembertian.
Recalling the unlensed waveform, one has
\begin{subequations}
    \begin{align}
        \hS0 &= \omega(r-t) ,\\
        \hA0 &= \frac{1}{r} .
    \end{align}
\end{subequations}
The $\mathcal{O}(U^{1})$ term gives
\begin{equation}\label{eq:S1}
    (\pt+\pr)\hS1 = -2 \omega U ,
\end{equation}
which has the solution
\begin{equation}
    \hS1 = -2\omega \int_{0}^{r} U(t-r+\rtld,\rtld,\theta,\varphi)d\rtld,
\end{equation}
where the potential is written in the spherical coordinates $U(t,\rv)=U(t,r,\theta,\varphi)$.
The $A^{(1)}$ term satisfies
\begin{equation}\label{eq:A1}
    (\pt + \pr) ( r A^{(1)} ) = (\pt + \pr) U - \frac{1}{2 \omega r^2} \nabla_\theta^2 S^{(1)} ,
\end{equation}
where $\nabla_\theta^2 \equiv  \partial_\theta^2 + \cot\theta\; \partial_\theta + \csc^2\theta\; \partial_\varphi^2$.
Recalling that $\gb_{ab,c}\sim \mathscr{R}^{-1} \ll \lambda^{-1}$, the ratio between $\partial(r A^{(1)})$ and $\partial \hS1$ can be written as $ \frac{\partial(r A^{(1)} )}{\partial \hS1} \sim \frac{\partial U}{\omega U} \sim \frac{\mathscr{R}^{-1} }{\lambda^{-1}} \ll 1 $.
Then, the lensed field $\phiL$ in the $L'$-plane is given by
\begin{equation}\label{eq:phi_Lp}
    \begin{aligned}
        \phiL(t,\rv)|_{L'} &= (\hA0 + \hA1) e^{i\hS0+i\hS1} = \hA0 e^{i\hS0} (1 + r \hA1) e^{i\hS1} \\
                           &  \simeq \hA0 e^{i\hS0+i\hS1} = \frac{1}{s'} e^{-i\omega[t-s'+\Psihat(t,s',\xiv')]} ,
    \end{aligned}
\end{equation}
with
\begin{equation}
    \Psihat(t,s',\xiv') \equiv 2 \int_{0}^{s'} U(t-s'+\rtld,\rtld,\theta,\varphi)d\rtld ,
\end{equation}
where $\xiv'$ is the coordinate in $L'$-plane, $\xiv'=(s'\sin\theta\cos\varphi,s'\sin\theta\sin\varphi)$.

Beyond the $L'$-plane, the wave equation becomes
\begin{equation}
    \Box \phiL=0 ,
\end{equation}
due to the condition (c).
Therefore, the lensed field $\phiL$ at the position of observer $O$ can be computed by the general form of Kirchhoff's theorem~\cite{1999prop.book.....B},
\begin{equation}
    \phiL(t,\rvo) = \frac{1}{\sqrt{2\pi}}\intff \phitldot(\rvo) e^{-i\omegatld t}d\omegatld ,
\end{equation}
with
\begin{equation}
    \phitldot(\rvo) = \frac{1}{4\pi}\int_{L'} d^{2}\xiv' \left\{ \phitldot \frac{\partial}{\partial z} \left( \frac{e^{i\omegatld l'}}{l'} \right)- \frac{e^{i\omegatld l'}}{l'} \frac{\partial \phitldot}{\partial z} \right\} .
\end{equation}

In this work, we focus on the gravitational lensing system in which the small-angle approximation is valid.
Then the equation above in the leading order becomes
\begin{equation}
    \phitldot(\rvo) = \frac{1}{4\pi}\int_{L'} d^{2}\xiv' \left\{ \phitldot \frac{e^{i\omegatld l'}}{l'} ( -i\omegatld -i \omega )\right\} ,
\end{equation}
which leads to
\begin{equation}
    \begin{aligned}
        \phiL(t,\rvo)
          &= \frac{1}{\sqrt{2\pi}}\intff \frac{1}{4\pi}\int_{L'} d^{2}\xiv' \left\{ \phitldot \frac{e^{i\omegatld l'}}{l'} ( -i\omegatld -i\omega )\right\} e^{-i\omegatld t}d\omegatld \\
          &= \frac{1}{4\pi}\int_{L'} d^{2}\xiv' ( \pt -i\omega ) \frac{1}{l'} \frac{1}{\sqrt{2\pi}}\intff \phitldot e^{-i\omegatld (t-l')} d\omegatld \\
          &= \frac{1}{4\pi}\int_{L'} d^{2}\xiv' ( \pt -i\omega ) \frac{\phiL(t-l')}{l'} \\
          &= \frac{1}{4\pi}\int_{L'} d^{2}\xiv' \frac{e^{-i\omega[t-l'-s'+\Psihat(t-l',s',\xiv')]}}{l's'} ( -2i\omega -i\omega\pt\Psihat ) ,
    \end{aligned}
\end{equation}
where the field solution in $L'$-plane, Eq.~\eqref{eq:phi_Lp}, is used in the last equality.

The amplification factor is a widely used quantity in the lensing theory, which is defined as the ratio between the lensed and unlensed waves at the observer position~\cite{1992grle.book.....S}
\begin{equation}
    F(t, \omega) \equiv \frac{\phiL(t,\rvo)}{\phi(t,\rvo)} = \frac{\phiL(t,\rvo)}{ \frac{1}{D_{OS}} e^{-i\omega(t-D_{OS})} } .
\end{equation}
Recalling the condition (a), one can identify the coordinates in $L'$-plane with coordinates in $L$-plane, which gives
\begin{equation}
    \begin{aligned}
        F(t, \omega)
        &= \frac{1}{4\pi}\int_{L} d^{2}\xiv \frac{D_{OS}}{ls} \exp\left(i\omega[l+s - D_{OS}-\Psihat(t-l,s,\xiv)]\right) ( -2i\omega -i\omega\pt\Psihat )\\
        &= (1+\frac{i}{2\omega}\pt) \left\{ \frac{\omega}{2\pi i}\int_{L} d^{2}\xiv \frac{D_{OS}}{ls} \exp\left(i\omega[l+s - D_{OS}-\Psihat(t-l,s,\xiv)]\right) \right\} .
    \end{aligned}
\end{equation}
Recalling the small-angle approximation, one has
\begin{subequations}
    \begin{align}
        l \simeq D_{OL} +\frac{1}{2}\xiv^{2}\frac{1}{D_{OL}} ,\\
        s \simeq D_{LS} +\frac{1}{2}\xiv^{2}\frac{1}{D_{LS}} ,
    \end{align}
\end{subequations}
and
\begin{equation}
    \Psihat(t-l,s,\xiv) \simeq 2 \int U(t-D_{OL}+z,\xiv,z)dz ,
\end{equation}
where the potential is written in the Cartesian coordinates with origin in the $L$-plane instead of the source, $U(t,\rv)=U(t,\xiv,z)$, and the integral has been approximated by collapsing the gravitational potential over the $z$-axis in the Cartesian coordinates.

Therefore, the amplification factor for the non-static gravitational lenses is given by
\begin{equation}\label{eq:Ft}
    \boxed{
        F(t, \omega) = (1+\frac{i}{2\omega}\pt) \left\{ \frac{\omega}{2\pi i} \frac{D_{OS}}{D_{OL}D_{LS}} \int_{L} d^{2}\xiv \exp\left(i\omega \left[\frac{1}{2}\xiv^{2} \frac{D_{OS}}{D_{OL}D_{LS}}-\Psi(t-D_{OL},\xiv) \right]\right) \right\}
    },
\end{equation}
with deflection potential
\begin{equation}\label{eq:Psi}
    \Psi(t,\xiv) \equiv 2 \int U(t+z,\xiv,z)dz .
\end{equation}
The Eq.~\eqref{eq:Ft} is the first main result of this work. It gives the amplification factor for the non-static gravitational lenses with the wave effect considered and is valid for non-static gravitational lenses that satisfy the metric Eq.~\eqref{eq:metric}.

There are two main terms in Eq.~\eqref{eq:Ft}.
The first term $\{\cdots\}$ is similar to the quasi-static approximation, in which the usual lensing solution is evaluated at each snapshot, effectively treating the lens as static at any given instant.
The second term $\frac{i}{2\omega}\pt\{\cdots\}$ arises from the next-order expansion of the wave equation in the presence of a non-static gravitational potential. It captures first-order corrections due to the genuine time dependence of the lens.
Formally, this term scales with $\omega^{-1}$. Thus, if the characteristic timescale of lens variation is much larger than the wave period, the lens remains effectively static over each wave cycle, and the time-derivative term is subdominant.

If the gravitational lenses are static, Eq.~\eqref{eq:Ft} recovers to the well-known result for static gravitational lenses~\cite{Nakamura:1999uwi,Takahashi:2003ix}
\begin{equation}\label{eq:F0}
    F_{0}(\omega) = \frac{\omega}{2\pi i} \frac{D_{OS}}{D_{OL}D_{LS}} \int_{L} d^{2}\xiv \exp\left(i\omega \left[\frac{1}{2}\xiv^{2} \frac{D_{OS}}{D_{OL}D_{LS}}-\Psi_{0}(\xiv) \right]\right) ,
\end{equation}
with
\begin{equation}
    \Psi_{0}(\xiv) \equiv 2 \int U(\cdot,\xiv,z)dz .
\end{equation}
Comparing Eq.~\eqref{eq:Ft} with Eq.~\eqref{eq:F0}, one can find there is an additional time dependence in the amplification factor which comes from the time dependence of the gravitational lenses.
A prominent feature of such a time-dependent amplification factor is that it can modulate the waveform of GWs, which will be discussed in detail in the following sections.

Usually, it's convenient to write the amplification factor as a function of dimensionless quantities.
Defining the following dimensionless quantities
\begin{subequations}
    \begin{align}
        \xv &\equiv (\xiv-\xivLb)/\xis, \\
        \yv &\equiv -\xivLb/\xis, \\
        w &\equiv \omega \ts, \\
        \tau &\equiv t/\ts, \\
        \psi &\equiv \Psi/\ts,
    \end{align}
\end{subequations}
where $\xis$ is an arbitrary normalization constant of length scale,
$\ts \equiv \frac{D_{OS}}{D_{OL}D_{LS}} \xis^{2}$ is the corresponding time scale,
and $\xivLb$ is a constant vector denoting the characteristic position of lenses.
The amplification factor can be rewritten as

\begin{equation}\label{eq:FtDimensionless}
    F(t, \omega) = (1+\frac{i}{2w}\ptau) \left\{ \frac{w}{2\pi i} \int_{L} d^{2}\xv \exp\left(iw \left[\frac{1}{2}|\xv-\yv|^{2} -\psi(\tau,\xv) \right]\right) \right\} .
\end{equation}

\subsection{Geometric optics limit}

When the frequency of GWs is very large, the amplification factor can be calculated by the stationary phase approximation.
Defining the dimensionless time delay as
\begin{equation}
    T(\xv; \yv,\tau) \equiv \frac{1}{2}|\xv-\yv|^{2} -\psi(\tau,\xv) .
\end{equation}
The stationary points of $T(\xv; \yv,\tau)$ are determined by
\begin{equation}
    \partial_{\xv} T(\xv; \yv,\tau) = 0.
\end{equation}
This stationary point $\xv_j(\yv,\tau)$ corresponds to the position of $j$-th image in geometric optics.
The time dependence of $\xv_j(\yv,\tau)$ means that the image position could vary with time.
Expanding $T(\xv; \yv,\tau)$ around the $j$-th image position $\xv_j(\yv,\tau)$, one has
\begin{equation}
    T(\xv; \yv,\tau)= T(\xv_j)+ \frac{1}{2} \sum_{a b} \bar{x}_a \bar{x}_b \partial_a \partial_b T(\xv_j) + \frac{1}{6} \sum_{abc} \bar{x}_a \bar{x}_b \bar{x}_c \partial_a \partial_b \partial_c T(\xv_j) + \cdots,
\end{equation}
where $\bar{\xv} = \xv - \xv_j$ and the indices $\{a,b,c, \cdots\}$ run from 1 to 2.
When the frequency $w$ is so large as to satisfy~\cite{Nakamura:1999uwi}
\begin{equation}
    w | \partial_{\xv}^2 T |^3 \gg | \partial_{\xv}^3 T |^2, \quad w | \partial_{\xv}^2 T |^2 \gg |\partial_{\xv}^4 T|, \quad \cdots ,
\end{equation}
the third and higher-order terms of this expansion can be neglected.
Then the amplification factor Eq.~\eqref{eq:FtDimensionless} becomes
\begin{equation}
    \boxed{
        F(t, \omega) = \sum_j F_{j}(t,\omega) = \sum_j (1+\frac{i}{2w}\ptau) \left\{ \Big| \mu\big(\xv_j(\yv,\tau)\big) \Big|^{1/2} \exp\Big( i w T\big(\xv_j(\yv,\tau)\big) - i \pi n_j \Big) \right\}
    },
\end{equation}
where
\begin{equation}
    \mu\big(\xv_j(\yv,\tau)\big) = 1 /\det \left[ \partial_a \partial_b T\big(\xv_j(\yv,\tau)\big) \right],
\end{equation}
and $n_j = 0, 1/2, 1$ when $\xv_j(\yv,\tau)$ is a minimum, saddle, maximum point of $T(\xv; \yv,\tau)$, respectively.

The $j$-th image of lensed monochromatic waves with frequency $\omega$ is given by
\begin{equation}
    \phiL_{j}(t, \rv) = \phi(t,\rv) F_{j}(t,\omega) = \frac{1}{r} e^{-i\omega(t-r)} F_{j}(t,\omega) ,
\end{equation}
where
\begin{equation}
    \begin{aligned}
    &F_{j}(t, \omega) \\
    &= \exp\Big( i w T\big(\xv_j(\yv,\tau)\big) - i \pi n_j \Big) \Big| \mu\big(\xv_j(\yv,\tau)\big) \Big|^{1/2}  \left( 1-\frac{1}{2}\ptau T\big(\xv_j(\yv,\tau)\big) +\frac{i}{4w} \ptau \ln \Big| \mu\big(\xv_j(\yv,\tau)\big) \Big| \right).
    \end{aligned}
\end{equation}
The magnification of the $j$-th image is
\begin{equation}
    \Big| \mu\big(\xv_j(\yv,\tau)\big) \Big|^{1/2} \left| 1-\frac{1}{2}\ptau T\big(\xv_j(\yv,\tau)\big) +\frac{i}{4w} \ptau \ln \Big| \mu\big(\xv_j(\yv,\tau)\big) \Big| \right|.
\end{equation}
Compared with static lens cases, the image magnification in non-static lens cases is time-dependent.
The first term, $\big| \mu\big(\xv_j(\yv,\tau)\big) \big|^{1/2}$, describes the image magnification in the quasi-static approximation, which is commonly used for microlensing light curves involving lens motion.
The second term, $\big| 1-\frac{1}{2}\ptau T + \frac{i}{4w} \ptau \ln|\mu| \big|$, represents the time-derivative correction. This correction can produce measurable effects when the quasi-static approximation breaks down, for example in cases of rapid lens motion.

Due to the time dependence of lenses, the image is no longer monochromatic, and it contains all the frequencies $w'$ from
\begin{equation}
    \int d\tau e^{i (w'-w) \tau} \exp\Big( i w T\big(\xv_j(\yv,\tau)\big) \Big) \Big| \mu\big(\xv_j(\yv,\tau)\big) \Big|^{1/2}  \left( 1-\frac{1}{2}\ptau T\big(\xv_j(\yv,\tau)\big) +\frac{i}{4w} \ptau \ln \Big| \mu\big(\xv_j(\yv,\tau)\big) \Big| \right) .
\end{equation}
When the time dependence of magnification is negligible, one can expand $T\big(\xv_j(\yv,\tau)\big)$ in $\tau$ up to the linear order around a reference time $\tau_0$, which leads to
\begin{equation}
    \propto \int d\tau e^{ i \tau \left(w'- w \left[ 1- \ptau T\big(\xv_j(\yv,\tau_0)\big) \right] \right) }.
\end{equation}
Then the image is approximately monochromatic around reference time $\tau_0$ with Doppler-shifted frequency
\begin{equation}
    w \left[ 1- \ptau T\big(\xv_j(\yv,\tau_0)\big) \right].
\end{equation}

\subsection{Cosmological situation}

The results above can be generalized to the cosmological situation.
We consider a spatially flat FLRW cosmology.
Since the gravitational potential is only effective in the neighborhood of the lenses, the background metric can be described by
\begin{equation}
    ds^{2} = a^2(\eta) \big\{ -[1+2U(\eta,\rv)]d\eta^{2} + [1-2U(\eta,\rv)]d\rv^{2} \big\} ,
\end{equation}
where $a(\eta)$ is the scale factor, $\eta$ is the conformal time and $ \rv $ is the comoving coordinate.
We set the present scale factor to be unity following the usual convention.

In this work, we focus on the GWs whose wavelengths are much smaller than the horizon scale.
Therefore, the spherical outgoing waves from the source have the waveform~\cite{maggiore2008gravitational}
\begin{equation}
    \phi(t(\eta),\rv) = \frac{1}{a(\eta)r} e^{-i\omega(\eta-r)}.
\end{equation}
Similarly, one can derive the amplification factor
\begin{equation}
    F(t(\eta), \omega) = (1+\frac{i}{2\omega}\peta) \left\{ \frac{\omega}{2\pi i} \frac{D_{OS}}{D_{OL}D_{LS}} \int_{L} d^{2}\xiv \exp\left(i\omega \left[\frac{1}{2}\xiv^{2} \frac{D_{OS}}{D_{OL}D_{LS}}-\Psi(\eta-D_{OL},\xiv) \right]\right) \right\}
    ,
\end{equation}
where $\xiv$ is the comoving coordinate and $D_{OS}, D_{OL}, D_{LS}$ are comoving distances.

The timescale of GWs passing through the lenses is much shorter than the cosmological timescale.
In the neighborhood of lenses when GWs pass through, one can define local coordinates as follows
\begin{equation}
    \begin{aligned}
        \tf &\equiv a_L\eta, \\
        \zetav &\equiv a_L\xiv, \\
        \zf &\equiv a_Lz,
    \end{aligned}
\end{equation}
where $a_L \equiv a(\eta_L) \equiv 1/(1+Z_L)$, $\eta_L$ is the conformal time when GWs pass through the lenses and $Z_L$ is the corresponding redshift.
In such local coordinates, the background metric can be expressed as
\begin{equation}
    ds^{2} =  -[1+2U(\tf,\zetav,\zf)]d\tf^{2} + [1-2U(\tf,\zetav,\zf)](d\zetav^{2}+d\zf^{2})  ,
\end{equation}
which has the same form as Eq.~\eqref{eq:metric}.
Similarly, one can define deflection potential in such local coordinates as
\begin{equation}
    \Upsilon(\tf,\zetav) \equiv 2 \int U(\tf+\zf,\zetav,\zf)d\zf ,
\end{equation}
which is related to $\Psi$ by
\begin{equation}
    \Psi(\eta-D_{OL},\xiv) = \frac{1}{a_L}\Upsilon(\tf-a_L D_{OL},\zetav).
\end{equation}

To simplify the results, one can use the angular diameter distance instead of comoving distance~\cite{Hogg:1999ad},
\begin{equation}
    \begin{aligned}
        d_{OL} &=a_L D_{OL} = D_{OL}/(1+Z_L), \\
        d_{OS} &=a_S D_{OS} = D_{OS}/(1+Z_S),\\
        d_{LS} &=a_S D_{LS} = D_{LS}/(1+Z_S),
    \end{aligned}
\end{equation}
where $Z_S$ is the redshift of the source.
Then the amplification factor can be expressed as
\begin{equation}\label{eq:Ft-cosmo}
    F(t, \omega) = (1+\frac{i}{2\omegaZL}\ptf) \left\{ \frac{\omegaZL}{2\pi i} \frac{d_{OS}}{d_{OL}d_{LS}} \int_{L} d^{2}\zetav  \exp\left(i\omegaZL \left[\frac{1}{2}\zetav^{2}  \frac{d_{OS}}{d_{OL}d_{LS}}-\Upsilon(\tf-d_{OL},\zetav) \right]\right) \right\} ,
\end{equation}
where $\omegaZL \equiv \omega(1+Z_L)$.
Since $a(\eta)$ evolves appreciably only on a cosmological timescale, once we fix $\eta=t$ at one moment in the present epoch, we have $\eta=t$, with exceedingly good accuracy over the whole timescale relevant for GW observation at a detector.
Therefore, one can normalize conformal time so that $t \simeq \eta$ at the present epoch.
Then, in order to calculate the amplification factor at the present epoch, one can simply do the calculations in the local coordinates of lenses following Eq.~\eqref{eq:Ft-cosmo}, and then transfer the results to cosmic time by $\tf \simeq a_L t = t/(1+Z_L)$.
If the gravitational lenses are static, the result above reduces to the well-known result for static gravitational lenses in the cosmological situation.

\section{Typical examples of non-static gravitational lenses}\label{sec:typical}

In reality, many astrophysical lenses are dynamic, with time-varying mass distributions due to phenomena such as moving stars, orbiting galaxies, or varying density distributions in galaxy clusters.
In this section, we consider two typical examples of non-static gravitational lenses: uniformly moving lenses and orbiting binary lenses~\cite{1986Natur.324..349G, Qiu:2022dya}.
For demonstration, we focus on the low redshift cases.

\subsection{Uniformly moving lenses}

\begin{figure}[htbp]
    \centering
    \includegraphics[]{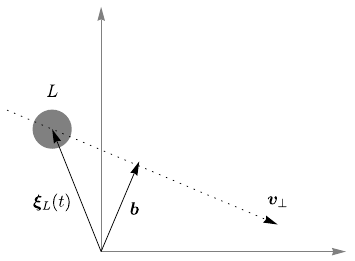}
    \caption{Diagram for the uniformly moving gravitational lenses in the lens plane.
    }
    \label{fig:uniformlyMoving}
\end{figure}

Consider gravitational lenses moving uniformly in the lens plane as shown in Fig.~\ref{fig:uniformlyMoving}.
Denoting the position of the lens as $\xivL$, one has
\begin{equation}
    \xivL(t) = \bv + \vt(t-t_{0}).
\end{equation}
Here, $\bv$ is the impact vector indicating the minimum distance between the lens trajectory and the line of sight, $t_{0}$ is the time of closest approach, and $\vt$ is the component of the lens velocity that is perpendicular to the line of sight.

We focus on the nonrelativistic moving lenses in this work.
Recalling the thin-lens approximation, the density distribution of lenses can be written as~\cite{1992grle.book.....S}
\begin{equation}
    \rho(t,\xiv,z) \propto \delta(z).
\end{equation}
Since the time derivative of $U$ is small, the deflection potential can be expressed as
\begin{equation}
    \Psi(t,\xiv) = \Psi_{0}(\xiv; \xivL(t)),
\end{equation}
where $\Psi_{0}(\xiv; \xivL)$ is the deflection potential for static lenses at position $\xivL$.
Therefore, the amplification factor for uniformly moving gravitational lenses is given by
\begin{equation}
    \begin{aligned}
        &F(t, \omega) \\
        &= (1+\frac{i}{2\omega}\pt) \left\{ \frac{\omega}{2\pi i} \frac{D_{OS}}{D_{OL}D_{LS}} \int_{L} d^{2}\xiv \exp\left(i\omega \left[\frac{1}{2}\xiv^{2} \frac{D_{OS}}{D_{OL}D_{LS}}-\Psi_{0}(\xiv; \xivL(t-D_{OL})) \right]\right) \right\}  \\
        &= (1+\frac{i}{2\omega}\pt) F_{0}\Big(\omega, \xivL(t-D_{OL})\Big) .
    \end{aligned}
\end{equation}

As an example, we consider a point-mass gravitational lens.
Set the normalization constant to be the Einstein radius, $\xis = \RE=\sqrt{4\ML D_{OL}D_{LS}/D_{OS}}$, where $\ML$ is the lens mass.
For a static point-mass gravitational lens, its amplification factor is known to be~\cite{Nakamura:1999uwi,Takahashi:2003ix}
\begin{equation}
    F_{0}(w,\yv) = K(w) \pFq{1}{1}(\frac{iw}{2},1;\frac{iw}{2}y^{2}),
\end{equation}
with
\begin{equation}
    K(w) \equiv \exp(\frac{\pi w}{4} + \frac{iw}{2} \ln \frac{w}{2}) ~\Gamma(1-\frac{iw}{2}),
\end{equation}
where $\yv \equiv -\xivLb/\xis=-\bv/\xis$, $y \equiv |\yv|$, $\pFq{1}{1}$ is the confluent hypergeometric function, and $\Gamma$ is the Gamma function.
Thus, the amplification factor for a uniformly moving point-mass gravitational lens is given by
\begin{equation}
    F(t, \omega) = (1+\frac{i}{2\omega}\pt) F_{0}\Big(\omega, -\left[ \bv + \vt(t-D_{OL}-t_{0}) \right]/\xis \Big) .
\end{equation}
Rewriting it as a function of dimensionless quantities, it becomes
\begin{equation}\label{eq:Ft_moving}
    F(t, \omega)= K(w)~ (1+\frac{i}{2w}\ptau) \pFq{1}{1}(\frac{iw}{2},1;\frac{iw}{2}[y^{2}+(\tau-\tauL)^{2}/\tauE^{2}]) ,
\end{equation}
where
\begin{equation}
    \begin{aligned}
        \tauE &\equiv \tE/\ts, \quad \tE \equiv \xis/|\vt| = \RE/|\vt| ,\\
        \tauL &\equiv \tL/\ts, \quad \tL \equiv t_{0} + D_{OL} .
    \end{aligned}
\end{equation}

\begin{figure}[htbp]
    \centering
    \includegraphics[width=.32\textwidth]{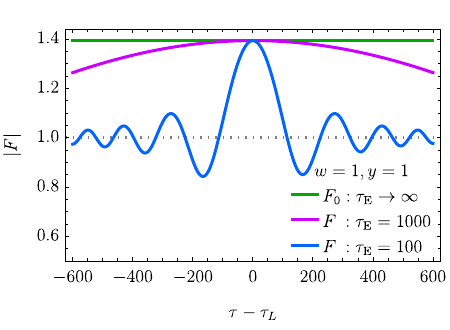}
    \includegraphics[width=.32\textwidth]{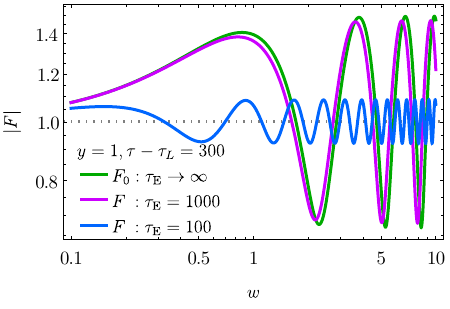}
    \includegraphics[width=.32\textwidth]{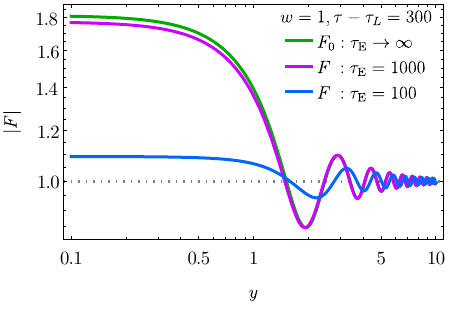}
    \caption{Amplification factor as a function of $\tau$, $w$, $y$ with different parameters.}
    \label{fig:moving_F}
\end{figure}

Fig.~\ref{fig:moving_F} shows $|F|$ as a function of $\tau$, $w$, $y$ with different parameters respectively.
The left panel shows explicitly that $|F|$ varies with time which is significantly different from the constant $|F_0|$ of the static gravitational lenses.
The middle and right panels show that $|F|$ and $|F_0|$ as functions of $w$ and $y$ have similar oscillating behavior.

To illustrate the effect of the time-derivative term, we decompose Eq.~\eqref{eq:Ft_moving} as
\begin{equation}
    F(t, \omega) = F_{\mathrm{qs}}(t, \omega) + F_{\mathrm{pt}}(t, \omega),
\end{equation}
where
\begin{align}
    F_{\mathrm{qs}}(t, \omega) &= K(w)~  \pFq{1}{1}(\frac{iw}{2},1;\frac{iw}{2}[y^{2}+(\tau-\tauL)^{2}/\tauE^{2}]) , \\
    F_{\mathrm{pt}}(t, \omega) &= K(w)~ \frac{i}{2w}\ptau \left\{ \pFq{1}{1}(\frac{iw}{2},1;\frac{iw}{2}[y^{2}+(\tau-\tauL)^{2}/\tauE^{2}]) \right\}.
\end{align}
Here, $F_{\mathrm{qs}}(t, \omega)$ corresponds to the quasi-static approximation, while $F_{\mathrm{pt}}(t, \omega)$ captures the time-derivative correction.

\begin{figure}[htbp]
    \centering
    \includegraphics[width=0.32\textwidth]{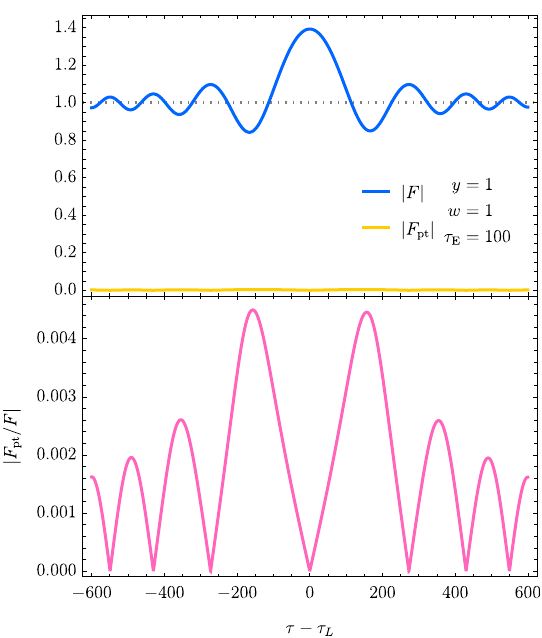}
    \includegraphics[width=0.32\textwidth]{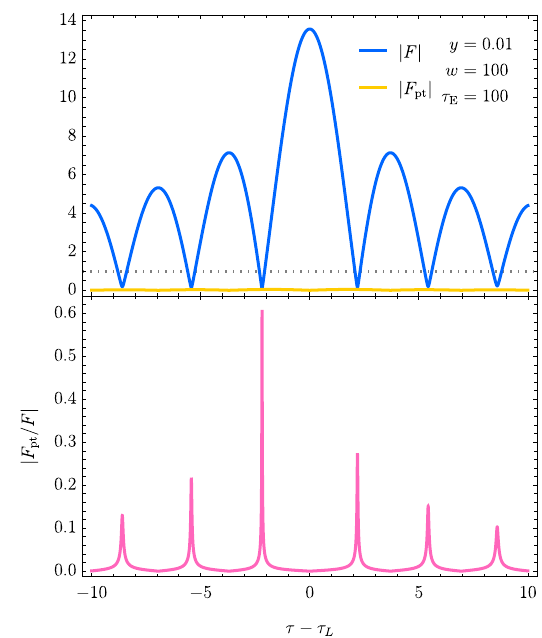}
    \includegraphics[width=0.32\textwidth]{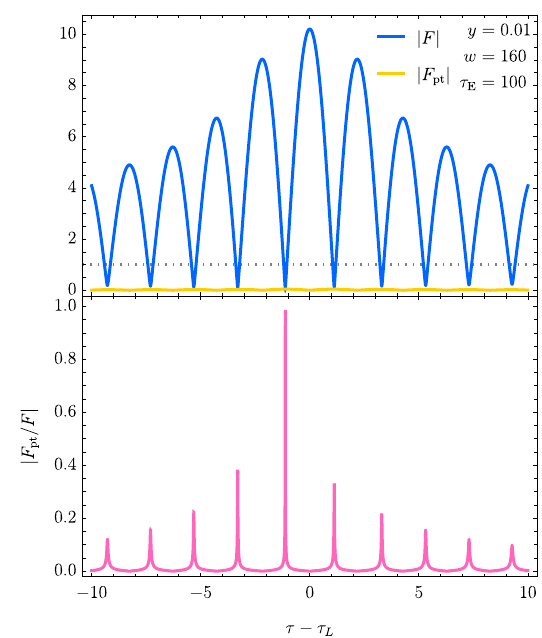}
    \caption{(\emph{Left}) Fractional difference $\left| F_{\mathrm{pt}}/F \right|$ as a function of $\tau$ for $\tauE=100$, $w=1$, and $y=1$. (\emph{Middle, Right}) Examples showing higher $w$ and smaller $y$ can make the time-derivative term more significant at certain intervals.
    }
    \label{fig:FptF}
\end{figure}

In the left panel of Fig.~\ref{fig:FptF}, we show the fractional difference $ \left| F_{\mathrm{qs}}/F -1 \right| = \left| F_{\mathrm{pt}}/F \right| $ as a function of $\tau$ for $\tauE=100$, $w=1$, and $y=1$.
In this regime, the discrepancy is on the order of $10^{-3}$.
However, non-static lensing and wave-optics effects can cause the amplification factor to oscillate in time, and there exist parameter regimes where the time-derivative term can exceed that $10^{-3}$ scale.
Indeed, in the middle and right panels of Fig.~\ref{fig:FptF}, we show scenarios with larger $w$ and smaller $y$, where $\left| F_{\mathrm{pt}}/F \right|$ becomes more pronounced at certain intervals. This underscores that while the time-derivative term may be modest in many cases, it can temporarily dominate under specific conditions.

\subsection{Orbiting binary lenses}

\begin{figure}[htbp]
    \centering
    \includegraphics[]{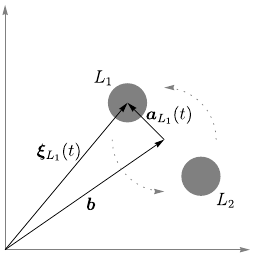}
    \caption{Diagram for the binary gravitational lenses in the lens plane.}
    \label{fig:binary}
\end{figure}

Consider the case of gravitational lenses consisting of binary objects.
For simplicity, we consider binary lenses with a circular orbit in the lens plane, as shown in Fig~\ref{fig:binary}.
In the Cartesian coordinates, the positions of the lenses are
\begin{subequations}
    \begin{align}
        \Big( \xivLi(t),\; 0 \Big) &= \Big( \bv+\avL1(t),\; 0 \Big) , \\
        \Big( \xivLii(t),\; 0 \Big) &= \Big( \bv+\avL2(t),\; 0 \Big) ,
    \end{align}
\end{subequations}
with
\begin{subequations}
    \begin{align}
        \avL1(t) &= +\frac{q}{1+q} R \Big( \cos[\omegaL (t-t_0)] ,\; \sin[\omegaL(t-t_0)] \Big) , \\
        \avL2(t) &= -\frac{1}{1+q} R \Big( \cos[\omegaL (t-t_0)] ,\; \sin[\omegaL(t-t_0)] \Big),
    \end{align}
\end{subequations}
where $\bv$ denotes the barycenter position of binary lenses in the lens plane, $R$ is the separation between two lensing objects which is much smaller than the distance to the source and observer such that the small-angle approximation is valid, $\omegaL$ is the angular frequency of the binary orbit, $t_0$ represents an initial time, and $ q \equiv M_{2}/M_{1} \leq 1$ is the mass ratio of the binary.

We focus on the nonrelativistic stage of binary lenses, i.e. $\omegaL R \ll 1$.
The lens potential of such binary lenses at $ (t, \xiv, z)$ is
\begin{equation}
    U(t, \xiv, z) = -\sum_{i=1,2} \frac{ M_i }{ \sqrt{ |\xiv - \xiv_{L_{i}}(t)|^2 + z^2 } } .
\end{equation}
Since the gravitational lensing effect is significant only in the region near the lenses, and the nonrelativistic motion of the lenses leads to a small time derivative of $U$, the deflection potential can be approximated as

\begin{equation}
    \begin{aligned}
        \Psi(t,\xiv) &= 2 \int U(t+z, \xiv,z) dz \\
                     &\simeq 2 \int U(t, \xiv,z) dz .
    \end{aligned}
\end{equation}
Hence, the amplification factor of binary lenses is
\begin{equation}
    F(t, \omega) = (1+\frac{i}{2\omega}\pt) \left\{ \frac{\omega}{2\pi i} \frac{D_{OS}}{D_{OL}D_{LS}} \int_{L} d^{2}\xiv \exp\left(i\omega \left[\frac{1}{2}\xiv^{2} \frac{D_{OS}}{D_{OL}D_{LS}}-\Psi(t-D_{OL},\xiv) \right]\right) \right\} ,
\end{equation}
with
\begin{equation}
    \Psi(t,\xiv) = \sum_{i=1,2} 4 M_i \ln | \xiv - \xiv_{L_i}(t ) | ,
\end{equation}
where the constant phase terms are neglected.

Set that $ \xi_* = \sqrt{4 (M_1 + M_2)D_{OL} D_{LS}/D_{OS}} $ and $\yv \equiv -\xivLb/\xis=-\bv/\xis$, the amplification factor of binary lenses can be rewritten as
\begin{equation}
    F(t, \omega) = (1+\frac{i}{2 w}\ptau) \left\{ \frac{w}{2\pi i} \int_{L} d^{2}\xv  \exp\left(i w \left[\frac{1}{2}|\xv-\yv|^{2} - \psi(\tau,\xv) \right]\right) \right\},
\end{equation}
where
\begin{equation}
    \psi(\tau,\xv) = \frac{1}{1+q} \ln | \xv - \xvLi(\tau)  | + \frac{q}{1+q} \ln | \xv -\xvLii (\tau)  |,
\end{equation}
and
\begin{subequations}
    \begin{align}
        \xvLi(\tau) &= \avL1(t-D_{OL})/\xis = +\frac{q}{1+q} \alpha \Big( \cos[\wL (\tau-\tauL)] ,\; \sin[\wL (\tau-\tauL)] \Big) , \\
        \xvLii(\tau) &= \avL2(t-D_{OL})/\xis = -\frac{1}{1+q} \alpha \Big( \cos[\wL (\tau-\tauL)] ,\; \sin[\wL(\tau-\tauL)] \Big),
    \end{align}
\end{subequations}
with dimensionless parameters
\begin{equation}
    \begin{aligned}
        \alpha &\equiv R/\xi_* , \\
        \wL &\equiv \omegaL \ts , \\
        \tauL &\equiv \tL/\ts, \quad \tL \equiv t_{0} + D_{OL} .
    \end{aligned}
\end{equation}

In the lens plane, one has $\xv = (x \cos\varphi,\; x \sin\varphi)$, and one can set $\yv = ( 0 ,\; y )$ without loss of generality.  Here $x \equiv |\xv|$ and $y \equiv |\yv|$.
Then the amplification factor of binary lenses can be rewritten as
\begin{equation}
    \begin{aligned}
        F(t, \omega) &= (1+\frac{i}{2 w}\ptau) \Bigg\{ \frac{w}{2\pi i} \int_{0}^{\infty} x\;dx \int_{0}^{2\pi} d \varphi\; \exp\Bigg(i w \Bigg[\frac{1}{2}(x^2 - 2 x y \sin\varphi + y^2)  \\
                     &- \frac{1}{2(1+q)} \ln \{x^2 - 2 x \frac{q}{1+q}\alpha \cos[\varphi - w_L (\tau - \tau_L)] + (\frac{q}{1+q}\alpha)^2\}  \\
                     &- \frac{q}{2(1+q)} \ln \{x^2 + 2 x \frac{1}{1+q}\alpha \cos[\varphi - w_L (\tau - \tau_L)] + (\frac{1}{1+q}\alpha)^2\} \Bigg]\Bigg) \Bigg\}.
    \end{aligned}
\end{equation}

\begin{figure}[htbp]
    \centering
    \includegraphics[width=.32\textwidth]{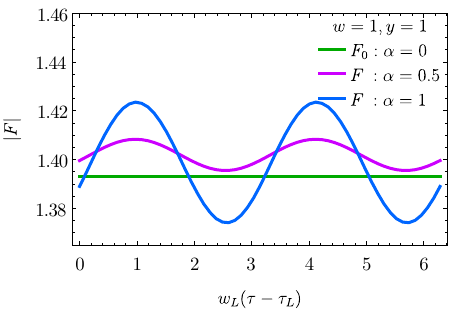}
    \includegraphics[width=.32\textwidth]{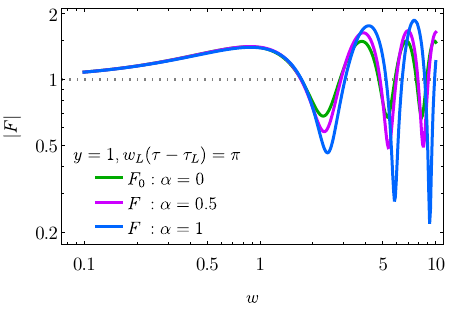}
    \includegraphics[width=.32\textwidth]{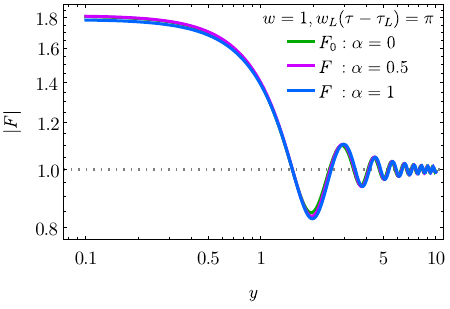}
    \caption{Amplification factor as a function of $\tau$, $w$, $y$ with different parameters, where $q=1$ is set for demonstration.}
    \label{fig:binary_F}
\end{figure}

Fig.~\ref{fig:binary_F} shows $|F|$ as a function of $\tau$, $w$, $y$ with different parameters respectively, where $q=1$ is set for demonstration.
The left panel shows that $|F|$ is periodic with period $\pi/w_{L}$ as a function of $\tau$.
This is because the orbiting lenses have equal mass ratio $q=1$ and period $\pi/\omegaL$.
The middle and right panels show that $|F|$ and $|F_0|$ as functions of $w$ and $y$ have similar oscillating behavior.

\section{Modulations of gravitational waves}\label{sec:modulation}

In this section, we consider different classes of GWs and show how they are modulated by non-static gravitational lenses.

\subsection{Monochromatic gravitational waves}

Firstly, we consider monochromatic GWs with frequency $\omegahat$.
Many sources in our universe can emit monochromatic GWs.
The spinning neutron stars can emit continuous GWs, which are long-lasting, quasi-monochromatic GWs with slowly varying intrinsic frequency. The frequency is in the band of current ground-based detectors, i.e. $\mathcal{O}(10)\Hz \sim \mathcal{O}(10^3)\Hz$~\cite{Jaranowski:1998qm,Prix:2009oha}.
QCD axions on stellar-mass black holes can emit GWs through superradiance. Such GWs, coming from axion annihilation to gravitons, are monochromatic and long-lived. Their frequencies are in the range of $\mathcal{O}(10^4)\Hz \sim \mathcal{O}(10^5)\Hz$, which depends on the axion mass~\cite{Arvanitaki:2009fg,Arvanitaki:2010sy,Arvanitaki:2012cn}.

When there are no gravitational lenses, the waveform and spectrum of GWs observed at the position $\rvo$ are
\begin{equation}
    \phi(t,\rvo) = A(\rvo) e^{-i\omegahat t},
\end{equation}
and
\begin{equation}
    \begin{aligned}
        \phitld(\omega,\rvo)
        = \frac{1}{\sqrt{2\pi}} \intff dt~e^{i\omega t} \phi(t,\rvo)
        = A(\rvo) \sqrt{2\pi}~\delta(\omega-\omegahat)
        .
    \end{aligned}
\end{equation}

After passing non-static gravitational lenses, the waveform and spectrum of GWs at the observer position $\rvo$ become
\begin{equation}
    \phi^{L}(t, \rvo)= A(\rvo) e^{-i\omegahat t} F(t, \omegahat) ,
\end{equation}
and
\begin{equation}
    \begin{aligned}
        \phitld^{L}(\omega,\rvo) = \frac{1}{\sqrt{2\pi}}\intff dt~e^{i\omega t} \phi^{L}(t, \rvo)
        = \frac{A(\rvo)}{\omegahat} \Htld(\omega,\omegahat)
        ,
    \end{aligned}
\end{equation}
where
\begin{equation}
    \Htld(\omega,\omegahat) \equiv \frac{\omegahat}{\sqrt{2\pi}} \intff dt~e^{i(\omega -\omegahat)t} F(t, \omegahat) .
\end{equation}

One can find that there are two characteristic effects: (i) time-varying modulations of the amplitude: comparing $|\phi^{L}(t,\rvo)|$ with $|\phi(t, \rvo)|$, the amplitude of GWs lensed by non-static gravitational lenses varies with time and is no longer a constant as that in the case without lenses; (ii) spectral broadening of monochromatic spectrum: comparing $|\phitld^{L}(\omega,\rvo)|$ with $|\phitld(\omega,\rvo)|$, the monochromatic waves become polychromatic after passing through non-static gravitational lenses.

For the uniformly moving point-mass gravitational lenses, recalling Eq.~\eqref{eq:Ft_moving}, one has
\begin{equation}
    \begin{aligned}
        &\Htld(\omega,\omegahat) \\
        &= \frac{\omegahat}{\sqrt{2\pi}} K(\wh) \intff \ts d\tau~e^{i(w-\wh)\tau} (1+\frac{i}{2\wh}\ptau) \pFq{1}{1}(\frac{i\wh}{2},1;\frac{i\wh}{2}[y^{2}+(\tau-\tauL)^{2}/\tauE^{2}]) \\
        &= \frac{\wh}{\sqrt{2\pi}} K(\wh) e^{i(w-\wh)\tauL} \frac{w+\wh}{2\wh} e^{\frac{i\wh}{2}y^{2}} \underset{n=0}{\overset{\infty }{\sum }} \frac{\left(-\frac{i \wh}{2} y^2\right)^n}{n!} \frac{i}{\pi} \sinh (\frac{\pi \wh}{2}) \Bigg\{ \\
        &~~~\Gamma(\frac{i\wh}{2}-\frac{1}{2}) \Gamma(1-\frac{i\wh}{2}) \frac{\left( 1-\frac{i\wh}{2} \right)_{n}}{\left( \frac{1}{2} \right)_{n}} \pFq{1}{1}\left(\frac{1}{2}-n;\frac{3}{2}-\frac{i \wh}{2};-\frac{i (w-\wh)^2 \tauE^2}{2 \wh}\right) \left( -\frac{i \wh}{2\tauE^2} \right)^{-\frac{1}{2}} \\
        &~~~+ 2\wh |w-\wh|^{-1+i\wh} \sinh (\frac{\pi \wh}{2}) \Gamma(\frac{i\wh}{2}) \Gamma(-i\wh) \\
        &~~~~~~~~~~~~~~~~~~~~~~~~~~~~ \times \pFq{1}{1}\left(\frac{i \wh}{2}-n;\frac{i \wh}{2}+\frac{1}{2};-\frac{i (w-\wh)^2 \tauE^2}{2 \wh}\right) \left(-\frac{i \wh}{2\tauE^2}\right)^{-\frac{i \wh}{2} } \Bigg\}
        ,
    \end{aligned}
\end{equation}
where $(a)_{n}\equiv\Gamma(a+n)/\Gamma(a)$ is the Pochhammer symbol.

\begin{figure}[htbp]
    \centering
    \includegraphics[width=0.48\textwidth]{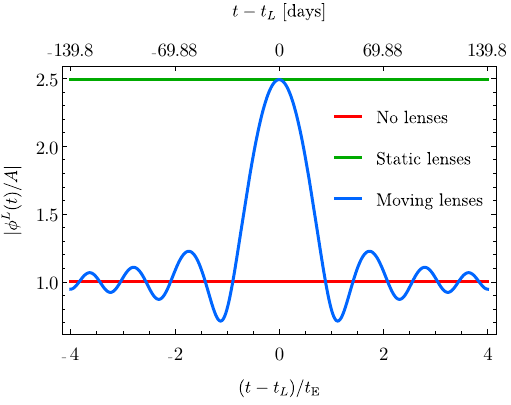}
    \includegraphics[width=0.48\textwidth]{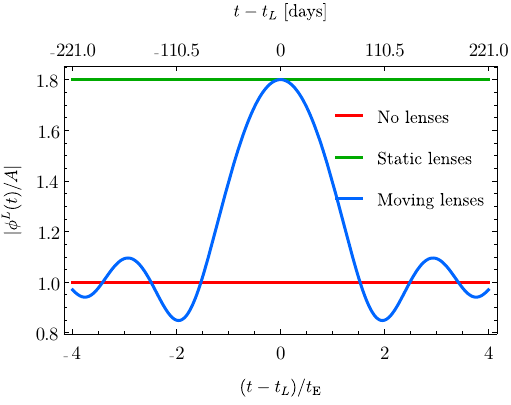}
    \includegraphics[width=0.48\textwidth]{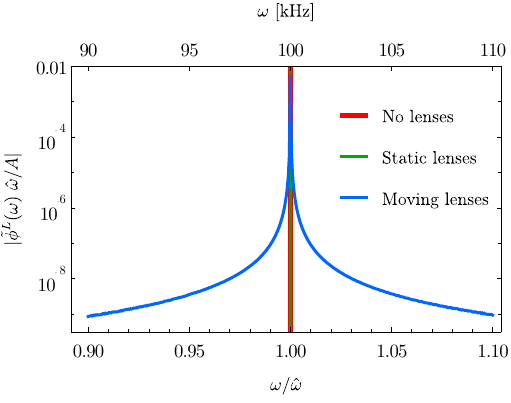}
    \includegraphics[width=0.48\textwidth]{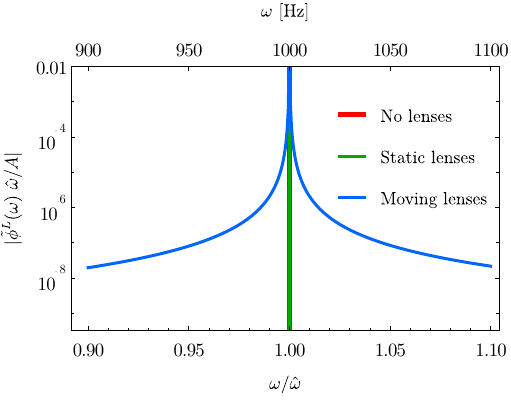}
    \caption{Normalized amplitude and spectrum of GWs lensed by uniformly moving point-mass gravitational lenses with $v_{L}=100 \mathrm{km/s}$ and $y=0$.
        \emph{Left:} $\omegahat=100\kHz$, $M_L=1\Msun$, $D_{OL}=D_{LS}=1\kpc$.
        \emph{Right:} $\omegahat=1000\Hz$, $M_L=50\Msun$, $D_{OL}=D_{LS}=50\pc$.
    }
    \label{fig:mono_moving}
\end{figure}

In figure~\ref{fig:mono_moving}, we show the normalized amplitude and spectrum of GWs lensed by uniformly moving point-mass gravitational lenses with $v_{L}=100 \mathrm{km/s}$ and $y=0$, which are denoted by blue lines.
The left panel demonstrates monochromatic GWs with $\omegahat=100\kHz$ passing through lens system with $M_L=1\Msun$, $D_{OL}=D_{LS}=1\kpc$.
The right panel demonstrates monochromatic GWs with $\omegahat=1000\Hz$ passing through lens system with $M_L=50\Msun$, $D_{OL}=D_{LS}=50\pc$.
The green lines describe the static lenses with the same parameters as the non-static lenses but $v_{L}=0$.
The red lines show the case without lenses.

Figure~\ref{fig:mono_moving} explicitly shows two characteristic effects caused by non-static gravitational lenses: time-varying amplitude modulations and spectral broadening of monochromatic spectra.
One can find that the amplitude of GWs oscillates with time. Besides, its envelope increases as the lens approaches the line of sight, and decreases as the lens leaves the line of sight. This is similar to the light curve of microlensing for electromagnetic waves.
Due to the velocity dependence, the lensed GWs by non-static gravitational lenses could be very useful in measuring the peculiar motion of celestial objects.

\begin{figure}[htbp]
    \centering
    \includegraphics[width=0.48\textwidth]{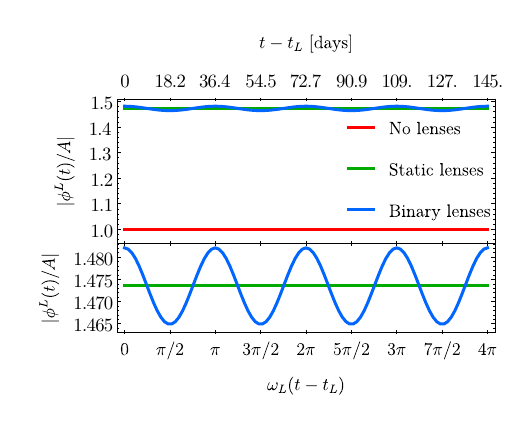}
    \includegraphics[width=0.48\textwidth]{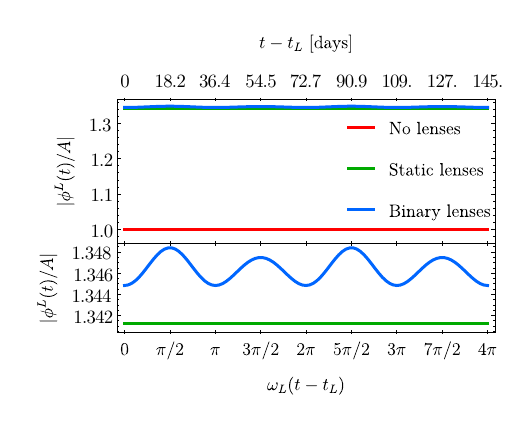}
    \caption{Normalized amplitude of GWs lensed by orbiting binary gravitational lenses with $\omegaL=10^{-6}\Hz$ and $y=1$.
        \emph{Left:} $\omegahat=100\kHz$, $M_{L_{1}}=M_{L_{2}}=1\Msun$, $D_{OL}=D_{LS}=1\kpc$.
        \emph{Right:} $\omegahat=1000\Hz$, $M_{L_{1}}=50\Msun$, $M_{L_{2}}=10\Msun$, $D_{OL}=D_{LS}=50\pc$.
    }
    \label{fig:mono_binary}
\end{figure}

In figure~\ref{fig:mono_binary}, we show the normalized amplitude of GWs lensed by orbiting binary gravitational lenses with $\omegaL=10^{-6}\Hz$ and $y=1$, which are denoted by blue lines.
The left panel demonstrates monochromatic GWs with $\omegahat=100\kHz$ passing through lens system with $M_{L_{1}}=M_{L_{2}}=1\Msun$, $D_{OL}=D_{LS}=1\kpc$.
The right panel demonstrates monochromatic GWs with $\omegahat=1000\Hz$ passing through lens system with $M_{L_{1}}=50\Msun$, $M_{L_{2}}=10\Msun$, $D_{OL}=D_{LS}=50\pc$.
The green line shows the case of static lenses with the same parameters but $\alpha=0$.
The red lines show the case where no lenses exist.
One can find that the amplitude of GWs lensed by orbiting binary lenses is modulated with frequency $2\omegaL$ ($\omegaL$) when an equal (unequal) mass ratio binary system is considered.
Due to the amplitude modulations, the spectrum of such lensed GWs will be polychromatic with peak frequencies $\omegahat \pm 2\omegaL n$ ($\omegahat \pm \omegaL n$) for an equal (unequal) mass ratio binary system, where $n$ is an integer.

\subsection{Polychromatic gravitational waves}

Now we consider polychromatic waves $\phi(t,\rv)$. They can be taken as the superposition of monochromatic waves $\frac{1}{\sqrt{2\pi}}\phitld(\omega,\rv) e^{-i\omega t}$:
\begin{equation}
    \phi(t,\rv) = \intff d\omega~\frac{1}{\sqrt{2\pi}} \phitld(\omega,\rv)~e^{-i\omega t}.
\end{equation}
After passing through the non-static gravitational lenses, the monochromatic waves are lensed
\begin{equation}
    \frac{1}{\sqrt{2\pi}}\phitld(\omega,\rv)~e^{-i\omega t} \rightarrow \frac{1}{\sqrt{2\pi}}\phitld(\omega,\rv)~e^{-i\omega t}~F(t, \omega).
\end{equation}
Therefore, the lensed waves will be
\begin{equation}
    \begin{aligned}
        \phi^{L}(t,\rv)
        & = \intff d\omega~\frac{1}{\sqrt{2\pi}} \phitld(\omega,\rv)~e^{-i\omega t}~F(t, \omega)
        .
    \end{aligned}
\end{equation}

For demonstration, we consider polychromatic GWs from the inspiral of circular orbiting binary.
The waveform of $+$ polarization is given by~\cite{maggiore2008gravitational}
\begin{equation}
    \begin{aligned}
        h_{+}(t) & \equiv \Re \{ \phi_{+}(t) \} =\Re \{ A(t) e^{-i\Phi(t)} \} \\
                 & = \frac{1}{d} \frac{G\Mc}{c^{2}} \left( \frac{5G\Mc}{c^{3}(\tc-t)} \right)^{1/4} \frac{1+\cos^{2}\iota}{2} \cos\left[ -2 \left(\frac{5G\Mc}{c^{3}(\tc-t)}\right)^{-5/8} + \Phi_0 \right] ,
    \end{aligned}
\end{equation}
where $d$ stands for the luminosity distance to the source, $\Mc$ represents the chirp mass, $\iota$ denotes the angle between the orbital angular momentum axis of a binary and the direction to the detector, $\tc$ and $\Phi_0$ are the time and phase at the coalescence time.

For the case we consider, the amplitude $A(t)$ of the GWs varies much more slowly than the phase $\Phi(t)$, and the instantaneous frequency $\dot{\Phi}(t)$ changes very slowly.
Concretely, this means $ \pt^n A(t) \ll \dot{\Phi}(t)^n$ for $n \ge 1$ and $\pt^n \Phi(t) \ll \dot{\Phi}(t)^n$ for $n>1$.
Therefore, one has
\begin{equation}
    \begin{aligned}
        \phi_{+}^L(t)
        & = \intff d\omega~\frac{1}{\sqrt{2\pi}} \phitld_{+}(\omega)~e^{-i\omega t}~(1+\frac{i}{2\omega}\pt) \left\{ \frac{\omega}{D}  \int_{L} d^{2}\xiv \exp\left[i\omega \Psitot(\xiv,t) \right] \right\} \\
        & = \int_{L} d^{2}\xiv [1-\frac{1}{2}\pt\Psitot(\xiv,t)] \intff d\omega~\frac{1}{\sqrt{2\pi}} \phitld_{+}(\omega)~e^{-i\omega t}  \frac{\omega}{D} \sum_{n=0}^{\infty} \frac{1}{n!} [i\omega \Psitot(\xiv,t)]^n \\
        & = \int_{L} d^{2}\xiv [1-\frac{1}{2}\pt\Psitot(\xiv,t)] \frac{1}{D} \sum_{n=0}^{\infty} \frac{1}{n!}  [i\Psitot(\xiv,t)]^n (i\pt)^{n+1} \intff d\omega~\frac{1}{\sqrt{2\pi}} \phitld_{+}(\omega)~e^{-i\omega t}  \\
        & = \int_{L} d^{2}\xiv [1-\frac{1}{2}\pt\Psitot(\xiv,t)] \frac{1}{D} \sum_{n=0}^{\infty} \frac{1}{n!}  [i\Psitot(\xiv,t)]^n (i\pt)^{n+1} \{ A(t) e^{-i\Phi(t)} \} \\
        & \simeq \int_{L} d^{2}\xiv [1-\frac{1}{2}\pt\Psitot(\xiv,t)] \frac{1}{D} \sum_{n=0}^{\infty} \frac{1}{n!}  [i\Psitot(\xiv,t)]^n A(t) e^{-i\Phi(t)} [\dot{\Phi}(t)]^{n+1} \\
        & = A(t) e^{-i\Phi(t)} F\left(t,\dot{\Phi}(t)\right)
        ,
    \end{aligned}
\end{equation}
where $D \equiv 2\pi i D_{OL}D_{LS} / D_{OS}$ and $\Psitot(\xiv,t) \equiv \left[\frac{1}{2}\xiv^{2} \frac{D_{OS}}{D_{OL}D_{LS}}-\Psi(t-D_{OS},\xiv) \right]$ are defined to simplify the notation of $F(t, \omega)$ in Eq.~\eqref{eq:Ft}.

\begin{figure}[htbp]
    \centering
    \includegraphics[width=0.48\textwidth]{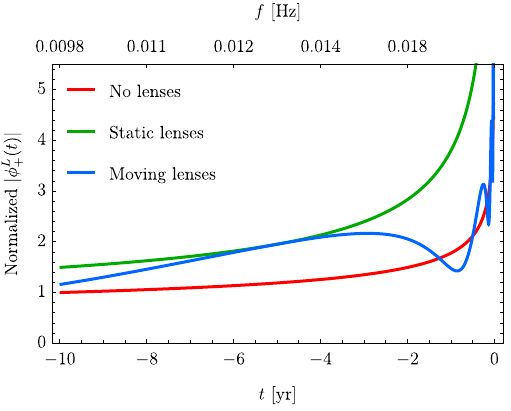}
    \includegraphics[width=0.48\textwidth]{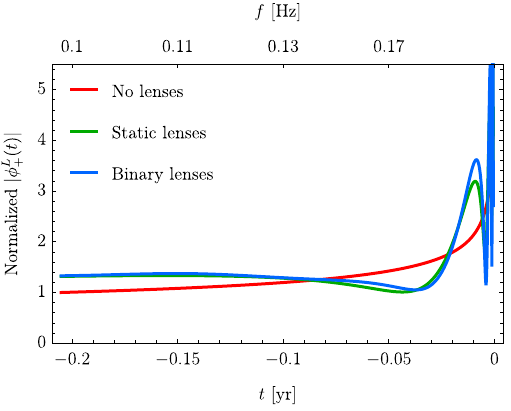}
    \caption{Normalized amplitude of lensed GWs originated from inspiral of circular orbiting binary at $D_{OS}=1\Mpc$.
        \emph{Left:} GWs from source with $M_{c,S}=40\Msun$ lensed by uniformly moving lenses with $\ML=5 \times 10^5 \Msun$, $D_{LS}=1\pc$, $\tL=-5\yr$, $v_{L}=100 \mathrm{km/s}$ and $y=0$.
        \emph{Right:} GWs from source with $M_{c,S}=10\Msun$ lensed by orbiting binary lenses with $M_{L_{1}}=M_{L_{2}}=5\times 10^4\Msun$, $D_{LS}=1\pc$, $\omegaL=10^{-6}\Hz$ and $y=1$.
    }
    \label{fig:poly}
\end{figure}

For simplicity, we set $\iota=0$, $\tc=0$, $\Phi_0=0$, and consider GW sources at $D_{OS}=1\Mpc$.
In Fig.~\ref{fig:poly}, we show the normalized amplitude of lensed GWs with blue lines.
The GWs in left panel are originated from sources with $M_{c,S}=40\Msun$, and lensed by uniformly moving lenses with $\ML=5 \times 10^5 \Msun$, $D_{LS}=1\pc$, $\tL=-5\yr$, $v_{L}=100 \mathrm{km/s}$ and $y=0$.
The GWs in right panel are originated from sources with $M_{c,S}=10\Msun$, and lensed by orbiting binary lenses with $M_{L_{1}}=M_{L_{2}}=5\times 10^4\Msun$, $D_{LS}=1\pc$, $\omegaL=10^{-6}\Hz$ and $y=1$.
In the left panel, compared with the green line which represents the GWs lensed by static lenses, the blue line oscillates obviously due to the time dependence of the moving lenses, which is similar to Fig.~\ref{fig:mono_moving}.
In the right panel, the blue line oscillates around the green line due to the periodic motion of orbiting binary lenses, which is also similar to Fig.~\ref{fig:mono_binary}.
Besides, one can find that the blue lines oscillate more and more rapidly with time since the instantaneous frequency is increasing.

\section{Conclusions}\label{sec:conclusion}

In this paper, we explored the gravitational lensing of GWs by non-static lenses, focusing on the effects of time-dependent mass distributions. While most existing studies of GW lensing assume static lenses, our work addresses the dynamic nature of realistic astrophysical lenses, such as moving stars and orbiting binaries. We developed a general theoretical framework that accounts for the time-dependent amplification of GWs and demonstrated how these non-static lenses can significantly modulate the observed signals.

Our analysis reveals that non-static gravitational lenses introduce unique modulating effects on GW signals, including time-varying amplitude modulations and spectral broadening. These effects were illustrated through two typical examples: uniformly moving lenses and orbiting binary lenses. For uniformly moving lenses, we found that the GW signal undergoes amplitude oscillations that depend on the lens's velocity and trajectory relative to the line of sight. In the case of orbiting binary lenses, the GW signals exhibit periodic modulations tied to the orbital motion, leading to distinctive spectral features. These modulations are not present in the case of static lenses and can provide new insights into the lensing objects and their dynamics.

Our findings have several important implications for GW astronomy and astrophysics. First, the ability to detect and characterize the modulations caused by non-static lenses can enhance our understanding of the mass distribution and dynamics of lensing objects like stars, black holes, or dark matter structures. This could lead to novel observational tests of general relativity and provide independent measurements of astrophysical parameters that are otherwise difficult to obtain.

Furthermore, the time-dependent amplification effects derived in this work offer a new tool for probing the peculiar motions and orbital dynamics of celestial objects. For instance, the detection of time-varying modulations in GW signals could be used to measure the velocities of lenses, offering a new method for studying stellar kinematics and the internal motions of galaxy clusters. This could significantly enrich our understanding of the evolution and interaction of cosmic structures.

Additionally, the unique signatures introduced by non-static lenses could help distinguish between different lensing scenarios and improve the interpretation of GW signals, especially in complex astrophysical environments. The presence of multiple lensed images with distinct time delays and modulations could also enhance the sensitivity of GW detectors, potentially enabling the discovery of otherwise undetectable sources.
As future GW observatories come online, the ability to recognize and interpret non-static lensing effects will become increasingly valuable.

Overall, our study underscores the need to account for time-dependent lensing effects in GW analyses and highlights the potential of non-static gravitational lenses as a powerful probe of the dynamic universe. Future work should focus on developing more sophisticated detection strategies for non-static lensing signatures and exploring a broader range of astrophysical scenarios. By integrating these dynamic effects into the broader framework of GW astronomy, we can unlock new observational windows into the universe’s most transient and energetic phenomena.

\section*{Acknowledgments}

We thank Jiuzhou Huang and Xi-Long Fan for the discussions.
XYY is supported in part by the KIAS Individual Grant No.~QP090702.
This work is supported in part by
the National Key Research and Development Program of China Grant  No. 2021YFC2203004, No. 2020YFC2201502 and No. 2021YFA0718304,
the National Natural Science Foundation of China Grants No. 12105344, No. 11821505, No. 11991052, No. 11947302, No. 12047503 and No.12235019.

\bibliographystyle{JHEP}
\bibliography{citeLib}

\end{document}